\documentclass[aps,prl,floatfix,twocolumn,footinbib,amsmath,amssymb,
  superscriptaddress]{revtex4-1}

\usepackage                             {epsfig}
\usepackage                             {natbib}
\usepackage                             {xspace}
\usepackage                             {amsmath}
\usepackage                             {graphicx}
\usepackage     [english]               {babel}
\usepackage                             {amsfonts}
\usepackage                             {amssymb}
\usepackage                             {latexsym}
\usepackage		                {color}
\usepackage       			{ulem}

\definecolor{darkgreen}{rgb}{0,.5,0}

\newcommand{\figref}[1]{Fig.~\ref{#1}}

\newcommand{\eref}[1]{(\ref{#1})}
\renewcommand{\eqref}[1]{Eq.~(\ref{#1})}
\newcommand{\ket}[1]{| #1 \rangle}

\newcommand{\braket}[1]{\langle #1 \rangle}
\begin{document}
\title{Coherent molecule formation in anharmonic potentials near
  confinement-induced resonances}

\author{S.\,Sala}
\affiliation{AG Moderne Optik, Institut f\"ur Physik,
  Humboldt-Universit\"at zu Berlin, Newtonstrasse 15, 12489 Berlin,
  Germany}
\email{ssala@physik.hu-berlin.de}

\author{G.\,Z\"urn}
\affiliation{Physikalisches Institut, Ruprecht-Karls-Universit\"at Heidelberg, Germany}
\affiliation{Max-Planck-Institut f\" ur Kernphysik, Saupfercheckweg 1, 69117 Heidelberg, Germany}

\author{T.\,Lompe}
\affiliation{Physikalisches Institut, Ruprecht-Karls-Universit\"at Heidelberg, Germany}
\affiliation{Max-Planck-Institut f\" ur Kernphysik, Saupfercheckweg 1, 69117 Heidelberg, Germany}
\affiliation{ExtreMe Matter Institute EMMI, GSI Helmholtzzentrum f\"ur Schwerionenforschung, Darmstadt, Germany}

\author{A.\,N.\,Wenz}
\affiliation{Physikalisches Institut, Ruprecht-Karls-Universit\"at Heidelberg, Germany}
\affiliation{Max-Planck-Institut f\" ur Kernphysik, Saupfercheckweg 1, 69117 Heidelberg, Germany}

\author{S.\,Murmann}
\affiliation{Physikalisches Institut, Ruprecht-Karls-Universit\"at Heidelberg, Germany}
\affiliation{Max-Planck-Institut f\" ur Kernphysik, Saupfercheckweg 1, 69117 Heidelberg, Germany}

\author{F.\,Serwane}
\affiliation{Physikalisches Institut, Ruprecht-Karls-Universit\"at Heidelberg, Germany}
\affiliation{Max-Planck-Institut f\" ur Kernphysik, Saupfercheckweg 1, 69117 Heidelberg, Germany}
\affiliation{ExtreMe Matter Institute EMMI, GSI Helmholtzzentrum f\"ur Schwerionenforschung, Darmstadt, Germany}
\affiliation{Now at the European Molecular Biology Laboratory, 69117 Heidelberg, Germany}

\author{S.\,Jochim}
\affiliation{Physikalisches Institut, Ruprecht-Karls-Universit\"at Heidelberg, Germany}
\affiliation{Max-Planck-Institut f\" ur Kernphysik, Saupfercheckweg 1, 69117 Heidelberg, Germany}
\affiliation{ExtreMe Matter Institute EMMI, GSI Helmholtzzentrum f\"ur Schwerionenforschung, Darmstadt, Germany}

\author{A.\,Saenz}
\affiliation{AG Moderne Optik, Institut f\"ur Physik,
  Humboldt-Universit\"at zu Berlin, Newtonstrasse 15, 12489 Berlin,
  Germany}

\date{\today}
\begin{abstract}
  We perform a theoretical and experimental study of a system of two
  ultracold atoms with tunable interaction in an elongated trapping
  potential. We show that the coupling of center-of-mass and relative
  motion due to an anharmonicity of the trapping potential leads to a
  coherent coupling of a state of an unbound atom pair and a molecule
  with a center of mass excitation. By performing the experiment with
  exactly two particles we exclude three-body losses and can therefore
  directly observe coherent molecule formation. We find quantitative
  agreement between our theory of inelastic confinement-induced
  resonances and the experimental results. This shows that the effects
  of center-of-mass to relative motion coupling can have a significant
  impact on the physics of quasi-1D quantum systems.

\end{abstract}
\maketitle
%
%
%

A key question in condensed matter physics is how the dimensionality
of a quantum system determines its physical properties.  Especially in
one dimension the increased role of quantum fluctuations leads to the
appearance of interesting phenomena which cannot be observed in
higher-dimensional systems.  This poses the interesting question of
how to experimentally realize such one-dimensional (1D) systems in a
three-dimensional (3D) world. This can be achieved by confining
particles in a strongly anisotropic potential whose lowest transversal
excitation is much larger than all other relevant energy scales of the
system.  In this case a 3D system can be mapped onto a true 1D system
obtaining an effective 1D coupling constant $g_{\text{1D}}$ which
depends on the 3D scattering length $a$ \cite{cold:olsh98}. In such
anisotropic confinement, ultracold atoms have been used to study,
e.g., the Tonks-Girardeau \cite{cold:pare04, cold:kino04, cold:kino05a}
and super-Tonks-Girardeau \cite{cold:hall09} gas as well as the
fundamental question of what constitutes an integrable quantum system
\cite{cold:kino06}.

Such experiments
\cite{cold:stoe06,cold:ospe06a,cold:hall09,cold:hall10a,cold:zurn12}
often
rely on the fact that it is possible to control the
effective 1D coupling strength $g_{\text{1D}}$ by tuning the scattering length
$a$ with a Feshbach resonance \cite{cold:chin10}. For a specific ratio of the
scattering length and the transversal confinement length $d_{\perp}$,
$g_{\text{1D}}$ diverges to $\pm \infty$ at a confinement-induced resonance
(CIR) \cite{cold:olsh98}. To distinguish these resonances in the elastic
scattering channel from the molecule-formation resonances we study in this
paper we will refer to them as \textit{elastic} CIRs.

A common experimental approach \cite{cold:hall10b} to characterize
such resonances has been to look for increased loss of atoms caused by
enhanced three-body recombination in the vicinity of the resonance.
However, this interpretation of the observed losses has been called
into question by a recent experiment which observed a splitting of
loss features under transversally anisotropic confinement
\cite{cold:hall10b}, although later theoretical works showed that no
such splitting of elastic CIR can occur
\cite{cold:peng10,cold:zhan11}. One proposed explanation for the
splitting is based on the fact that the trapping potentials used in
experiments are not perfectly harmonic. This leads to a coupling of
center-of-mass (COM) and relative (REL) motion
\cite{cold:pean05,cold:schn09,cold:kest10}, which in turn can lead to
a coupling of two atoms in the ground state of the trap to a weakly
bound molecular state with a COM excitation \cite{cold:sala12}
(further elaborated on in \cite{cold:peng11}). The occupation of the
bound state is only possible because the excess binding energy can be
transferred into COM excitation energy due the anharmonictiy of the
confining potential. This redistribution of binding energy to kinetic
energy is an inelastic process and thus we refer to these COM-REL
coupling resonances as \textit{inelastic} CIR.
	
In a many-body system losses at the inelastic CIR can be described as
a two-step process: First, two atoms coherently couple to the
COM-excited molecular state. Then, this molecule collides either with
another molecule or an unbound atom, which leads to a deexciation of
the molecule into a deeply bound state and subsequent loss of the
involved particles from the trap.

However, also different theoretical models have been developed to
explain the observed splitting of the loss features in
\cite{cold:hall10b}. These argue with enhanced three-body effects in
the vicinity of elastic CIR. One is based on multichannel effects
\cite{cold:mele11}, others \cite{cold:hall10b, cold:arim11} on a
Feshbach-type mechanism.  In a many-body system as used in
\cite{cold:hall10b} different loss mechanisms are in principle
possible and cannot be clearly distinguished by the experiment. A
straightforward, yet experimentally challenging solution to this
problem is to eliminate three-body effects by investigating a pure
2-body system. In this work we provide a direct experimental
confirmation of the theory developed in \cite{cold:sala12} by
performing a theoretical and experimental study of two $^6$Li atoms in
an elongated trapping potential with a slight ellipticity. Ab initio
calculations of the coupling strengths, the widths and the positions
of the coherent molecule formation at the inelastic CIR are found to
be in quantitative agreement with the experimental results.

%
%

To prepare a quasi-1D two-body system we follow the same preparation
scheme as described in \cite{cold:serw11,cold:zurn12}, which has a
fidelity of about 90\%. The two particles are trapped in the ground
state of a cigar-shaped potential with a mean transversal confinement
length of $d_\perp=0.486 \pm 0.006\, \mu$m and an aspect ratio of
about 10:1, which is well in the quasi-1D regime
\cite{cold:idzi06}. The shape and anharmonicity of the potential have
been characterized by precise measurements of the transition
frequencies for exciting a single particle into the first and second
excited level in the longitudinal and both transversal directions
\cite{cold:prl_sm}.

%
%

This two-body system is in absolute coordinates described by the
Hamiltonian
\begin{align}
  H(\mathbf{r_1}, \mathbf{r_2}) = T_1(\mathbf{r_1}) +
  T_2(\mathbf{r_2}) + V_1(\mathbf{r_1}) + V_2(\mathbf{r_2}) +
  U(|\mathbf{r_1}-\mathbf{r_1}|)
  \label{eq:hamil}
\end{align}
where $T_1$, $T_2$, $V_1$, $V_2$ denote the kinetic energies and
potential energies due to the trap of particles one and two,
respectively, and $U$ the interatomic interaction. It has been
demonstrated that sextic potentials, i.e. expansions of a $\sin^2$
optical-lattice potential up to order six,
\begin{align}
  V(\mathbf{r}) = \sum_{j=x,y,z} \frac{2}{45} V_{j} k_{j}^{6} j^{6} -
  \frac{1}{3} V_{j} k_{j}^{4} j^{4} + V_{j} k_{j}^{2} j^{2} ,
  \label{eq:sextic_absolute}
\end{align}
are well suited to describe anharmonicity induced COM-REL coupling in
single-well potentials \cite{cold:gris09}, like the one used in our
experiment.

The stationary Schr\"odinger equation for the Hamiltonian
(\ref{eq:hamil}) can be solved exactly by the computational approach
described in \cite{cold:gris11}. Herein, the interaction potential is
treated by a numerically given Born-Oppenheimer potential curve of a
$^6$Li system. The variation of the scattering length due to the
magnetic Feshbach resonance can be modeled computationally by
modifying the inner wall of the potential curve which effectively
changes the scattering length of the system to arbitrary values
\cite{cold:gris10}.

\begin{figure}[ht]
  \begin{centering}
    \includegraphics[width=0.45\textwidth]{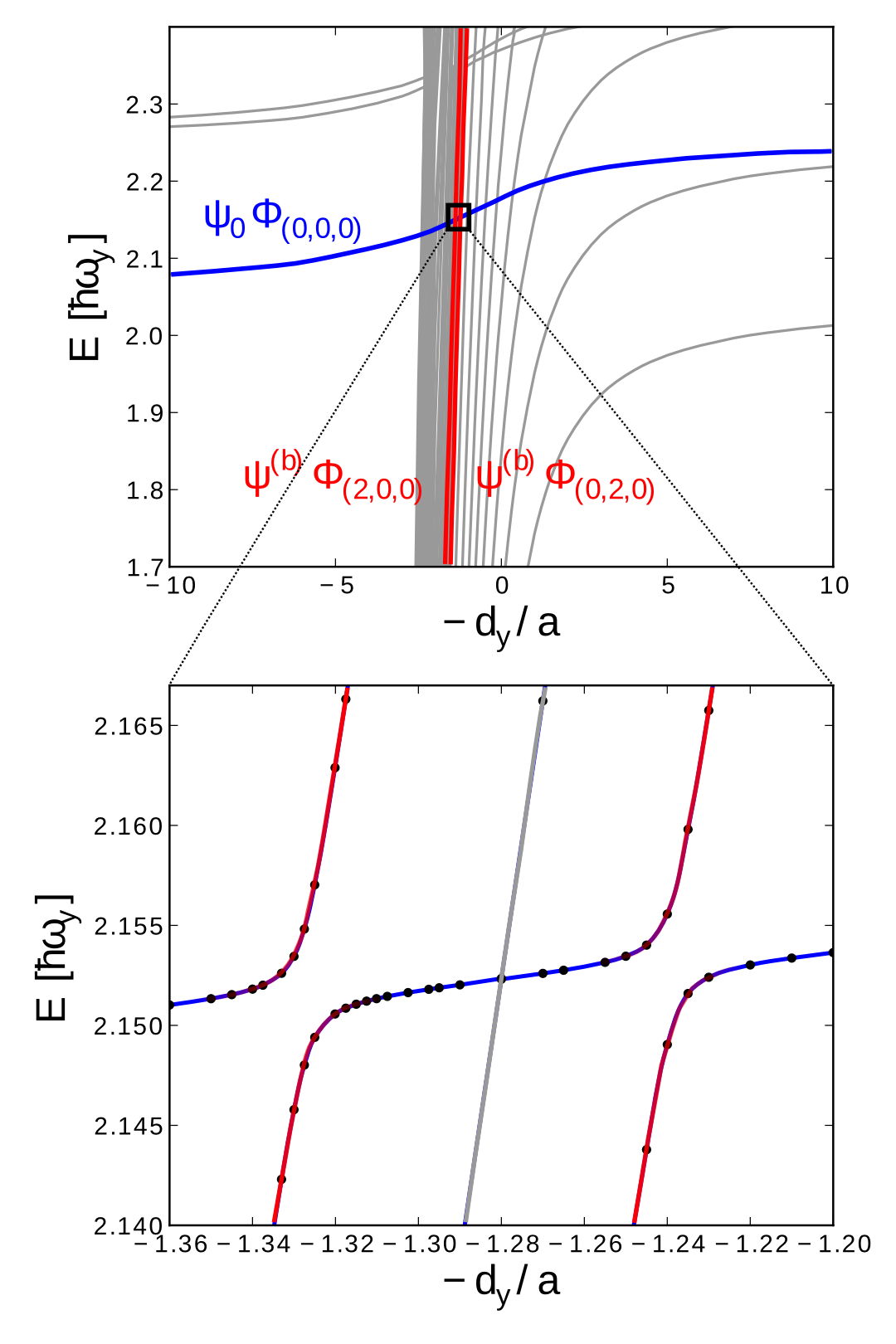}
    \caption{(color online) Eigenenergy spectrum of the Hamiltonian
      \eref{eq:6tic_Hamil_full} for $^6$Li atoms confined in a sextic
      trapping potential. In the upper part all states bending down to
      $-\infty$ are molecular states originating from the REL bound
      state $\psi^{(b)}$ with different COM excitations. The two bound
      states marked in red are the only ones which have a significant
      coupling to the repulsive state (blue). For the other states
      (gray) the coupling is negligible. The magnified part shows the
      avoided crossings responsible for the COM-REL resonances.  }
    \label{fig:full_spec}
  \end{centering}
\end{figure}

For a two-particle system it is convenient to transform the
Hamiltonian in REL and COM coordinates, $\mathbf{r}=
\mathbf{r_1}-\mathbf{r_2}$ and
$\mathbf{R}=\frac{1}{2}(\mathbf{r_1}+\mathbf{r_2})$, respectively,
\begin{align}
  H(\mathbf{r},\mathbf{R}) =\ & T_{\rm REL}(\mathbf{r}) + T_{\rm
    COM}(\mathbf{R}) + V_{\rm REL}(\mathbf{r})\nonumber\\ & + V_{\rm
    COM}(\mathbf{R}) + U_{\rm int}(r) + W(\mathbf{r}, \mathbf{R}).
\label{eq:6tic_Hamil_full}
\end{align}
$ V_{\text{REL}}(\mathbf{r})$ and $V_{\text{COM}}(\mathbf{R})$ are the
separable parts of the sextic potential \cite{cold:gris09}. Thus,
$W(\mathbf{r}, \mathbf{R})$ contains only the non-separable terms,
i.e.\ a polynomial in $r_j^2 R_j^2$, $r_j^2 R_j^4$ and $r_j^4 R_j^2$.
The potential parameters $V_j$ and $k_j$ are obtained by fitting the
eigenenergies of a single particle in a sextic potential to the
experimentally measured transition energies of a single particle in
the trap. The fit results are given in \cite{cold:prl_sm}. The eigenenergies
and wavefunctions of the Hamiltonian (\ref{eq:6tic_Hamil_full}) can
now be calculated via exact diagonalization for different values of
the s-wave scattering length.  A fully coupled spectrum is shown in
\figref{fig:full_spec}.

Relative motion bound states $\psi^{(b)}$ with COM excitation
$\Phi_{\mathbf{n}}$, $\mathbf{n}=(n_x,n_y,n_z)$ (i.e.\ states bending
down to negative infinity) cross with trap states, i.e.\ states whose
energy converges asymptotically to a constant value for $a \to
0^+$. In the absence of a trapping potential these states would lie in
the continuum. The system is initially in the lowest trap state, i.e.\
dominantly in the relative motion repulsive state $\psi_0$ and COM
ground state $\Phi_{(0,0,0)}$ (see blue state in
\figref{fig:full_spec}). Hence, it suffices to consider crossings with
this state. The coupling, and equivalently the size of the avoided
crossings, is described by the coupling matrix elements
\begin{align}
  W_\mathbf{n} = \braket{\psi^{(b)} \, \Phi_\mathbf{n} | W | \psi_0\,
    \Phi_{(0,0,0)}}.
  \label{eq:mat_elem}
\end{align}
In \cite{cold:sala12} it was demonstrated that in quasi 1D only the
lowest transversally COM excited bound states, $\ket{\psi^{(b)} \,
  \Phi_{(2,0,0)}}$ and $\ket{\psi^{(b)} \, \Phi_{0,2,0}}$ (see red
states in \figref{fig:full_spec}), couple significantly with the
lowest trap state.  Therefore, in \figref{fig:full_spec} only the
transversally excited bound states form significant avoided crossings
with the repulsive trap state.  Due to the transverse anisotropy of
the trap these crossings are non-degenerate which results in a
splitting of the resonances. Such a splitting was also observed in
\cite{cold:hall10b} in quantitative agreement with the positions of
the inelastic CIR \cite{cold:sala12}.

To demonstrate that the crossing states possess characteristics of a
bound and a trap state the mean radial density
\begin{align}
  \overline{r} = \int_0^{\infty} \mathrm{d}r \, r\, \rho(r).
\end{align}
was calculated. Here,
\begin{align}
  \rho(r) = r^2 \, \int \, \mathrm{d}V_\mathbf{R} \, \mathrm{d}\Omega_\mathbf{r} \,
  | \Psi(\mathbf{r}, \mathbf{R}) |^2
  \label{eq:raddens}
\end{align}
is the radial pair density where $\Psi(\mathbf{r}, \mathbf{R})$
denotes the full six-dimensional wavefunction of the system,
$\mathrm{d}V_\mathbf{R}$ is the COM volume element and
$\mathrm{d}\Omega_\mathbf{r}$ is the angular volume element of the REL
motion. At $d_y/a = 1.38$ the bound state has a mean radial distance
of $\overline{r} = 0.29\, d_{\perp}$, i.e.\ it is small compared to
the mean transversal confinement length.
This demonstrates the strong binding of the atoms. In the trap state,
the atoms possess a mean distance of $\overline{r} =1.19\,d_z =
3.06\,d_{\perp}$.  This mean distance which is of the order of the
longitudinal trap length $d_z = 1.25\,\mu$m is a consequence of the
elongated trap.

In the vicinity of the avoided crossing the system can be
approximately described as a two-level system because the other states
are energetically almost inaccessible. When the scattering length is
ramped non-adiabatically towards the crossing and stopped in the gap
region of the avoided crossing the system finds itself in a coherent
superposition of the two adiabatic states \cite{cold:syas07}, which in
our case are the bound state $\ket{\psi^{(b)} \, \Phi_\mathbf{n}}$ and
the repulsive trap state $\ket{\psi_0\, \Phi_{(0,0,0)}}$. Since both
states evolve with different phase a Rabi-oscillation between the
states occurs with the frequency
\begin{align}
  \Omega = \frac{1}{\hbar} \, \sqrt{W_\mathbf{n}^2 + \delta^2 }
\end{align}
which is a measure for the coupling strength for $\delta = (E_b
-E_t)/2 = 0$. Here, $W_\mathbf{n}$ is the coupling matrix element from
\eqref{eq:mat_elem} while $E_b$ and $E_t$ denote the energies of the
diabatic bound and trap states, respectively.

Experimentally this coherent superposition is realized by first
preparing two $^6$Li atoms in the ground state of the potential and
then increasing the scattering length $a$ by ramping up the magnetic
offset field non-adiabatically with a speed of $20\,$G/ms.  To locate
the molecule formation resonances the ramp is suddenly stopped at
different values of the magnetic offset field.  The population is
expected to oscillate between the unbound and the COM-excited
molecular state as a function of the Rabi frequency $\Omega$ which
depends on the magnetic field.

\begin{figure} [th!]
  \centering
  \includegraphics [width=0.45\textwidth]
  {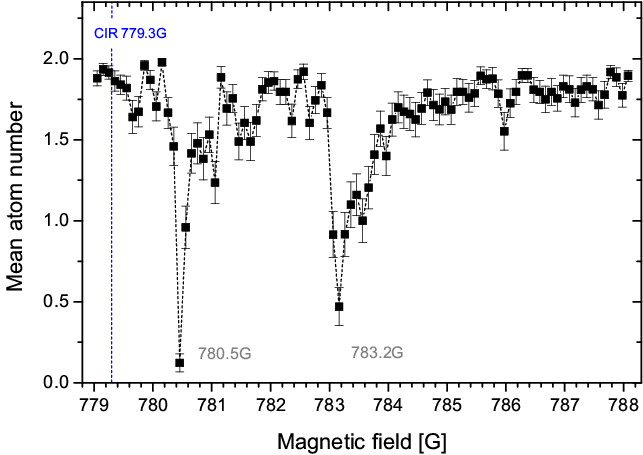}
  \caption{ Disappearance of particles in the repulsive non-bound
    state.  Due to the COM-REL motion coupling the particles in the
    non-bound state couple into a molecule and disappear when
    detecting the number of particles in the non-bound state. One
    observes two peaks indicating COM-REL motion coupling resonances
    involving two excited molecular states in $x$- and $y$-direction
    of the confinement. Each data point is the average of about 50
    individual measurements with discrete atom number. The blue dashed
    line indicates the position of the elastic CIR at $779.3\pm
    0.5\,$G calculated using the transversal confinement length
    $d_\perp$ and the calibration of the scattering length $a(B)$ of
    \cite{cold:zurn12b} as inputs for the theory of
    \cite{cold:berg03}. }
	\label{fig:figure2}
\end{figure}

\begin{figure} [th!]
\centering
\includegraphics [width=0.45\textwidth] {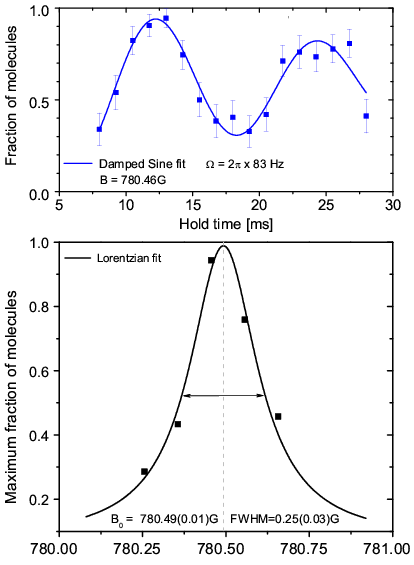}
\caption{Coherent dynamic of the COM-REL motion coupling. (a)
  Oscillation between the non-bound and the COM excited molecular
  state. From a sinusoidal fit we deduce the Rabi-frequency
  $\Omega$. (b) Maximum amplitude of the oscillation. The data points
  are extracted from measurements analog to figure a) at different
  magnetic offset fields.}
	\label{fig:figure3} 
\end{figure}

In a first experiment we wait for a fixed hold time of $12.5\,$ms
after stopping the ramp at different magnetic field values between
779\,G and 788\,G \footnote{ The duration of the hold time is such
  that it corresponds to a half-cycle (i.e. a $\pi$-pulse) of an
  expected Rabi-Frequency of $\Omega_0 = 2 \pi \times 80\,$Hz.}.
We then measure the number of free atoms remaining in the ground state
of the trap by ramping to a magnetic field of 523\,G where the
molecules are deeply bound and therefore not observed with our
detection scheme.  Thus, the mean number of molecules is given by
$N_{\text{mol}} = (N_0 - N_{\text{GS}}) / N_0$, where $N_0$ is the
mean number of atoms in the initial system and $N_{\text{GS}}$ is the
mean number of particles detected in the non-molecular ground state at
the end of the experiment.  To check whether the missing atoms indeed
end up in the molecular state we repeated the experiment but ramped
the magnetic field to a value of $900\,$G before measuring the number
of particles. At this magnetic field we are far above the elastic CIR
so that the molecules become weakly bound and the constituent
particles of the molecules can be detected with our detection
scheme. We found that there is no measurable change compared to the
initial particle number when measuring above the elastic CIR, which
excludes the presence of any significant loss channels in our system.
Figure \ref{fig:figure2} shows the detected number of particles in the
repulsive state depending on the magnetic offset field. As expected
from numerics, two peaks are observable which are identified as the
COM-REL motion coupling resonances created by the two molecular states
excited in
$x$- and $y$- direction of the anisotropic confinement. \\
To analyze the dynamics of the coupling we ramped to different values
of the magnetic offset field around the features shown in Fig.\
\ref{fig:figure2} and held the system for different hold times.  With
less than $10\%$ probability we detect only a single atom in the trap,
i.e. the two atoms are either free ($N=2$ detected) or bound to a
molecule ($N=0$). The few realizations with just a single atom
detected ($N=1$) are not considered in the
analysis. Fig.\,\ref{fig:figure3}\,a) shows the result of one of these
measurements.
The oscillation of the fraction of molecules shows that we have
created a coherent superposition of the molecular and the repulsive
state. By performing a sinusoidal fit to the data we can extract the
Rabi frequency $\Omega$ of the oscillation. The maximum amplitudes of
the oscillation for different magnetic fields are shown in figure
\ref{fig:figure3}\;b).  From a Lorentzian fit to the amplitude we can
extract the width (FWHM) of the coupling in terms of the magnetic
offset field. Table \ref{tab:com-rel-resonances} shows the width of
the coupling resonances determined from the measurement.
\begin{table}[h!]
\begin{center}
  \begin{tabular}{|c | c |c| c |c |c|c|}
    \hline
    COM  & \multicolumn{2}{|c|}{Position [G] }    & \multicolumn{2}{|c|}{FWHM[G]}  &  \multicolumn{2}{|c|}{$\Omega_{0}/ 2 \pi$\,[Hz]}  \\
    excitation & exp. & num. & exp. & num. &  exp. & num. \\ \hline
    $(0,2,0)$ & 780.5  & 776.01   &  0.25(0.03) \, \, & 0.35 &  83(2) \; \; & 64\\ 
    $(2,0,0)$ & 783.2  & 779.02   &  0.42(0.06)$^{(*)}$ &0.35  &  75(1) $^{(*)}$ & 69 \\ \hline
     \end{tabular}
\end{center}
\caption{Comparison between experiment and numerical calculation.\; $^{(*)}$\, See \cite{cold:prl_sm} for these measurements.
}
	\label{tab:com-rel-resonances}
\end{table}
The spacing between the two resonances, see Table
\ref{tab:com-rel-resonances}, is in agreement up to $0.3\,$G with the
numerical calculation. The absolute position of the experimental
resonances is shifted about $4.3\,$G compared to the theoretical
values. In view of the width of the elastic CIR\footnote{The width of
  the elastic CIR is mainly determined by the width of the Feshbach
  resonance \cite{cold:chin10}.} of $250\,$G this is a remarkable
accuracy. Moreover, except for the two COM-REL coupling resonances no
significant molecule formation was observed over the whole width of
the elastic CIR.

%
%

In conclusion, our results directly show that in a two-particle system
the COM-REL coupling allows for the coherent coupling of an unbound
atomic pair and a molecular state without a third particle being
present. Competing processes such as three-body recombination or
processes involving atoms in higher bands \cite{cold:mele11} are
excluded by our high preparation fidelity. Hence, the agreement
between the theoretical and experimental results gives a quantitative
confirmation of the theory of inelastic CIR \cite{cold:sala12}.
Furthermore, our results show that a molecule formation in a two-body
system is absent at the elastic CIR \footnote{ One should note that
  this is no contradiction to the description of the elastic CIR using
  a Feshbach type mechanism \cite{cold:berg03}.  In this picture a
  shifted bound state crosses the continuum threshold which one could
  expect to be responsible for molecule formation.  However, this
  shifted bound state of \cite{cold:berg03} is \textit{not} an
  eigenstate of the full Hamiltonian but of a modified one which
  results from a non-unitary transformation and hence does not couple
  to the repulsive trap state.  In contrast, the molecular states that
  are populated in the present work are bound states with COM
  excitation, i.e. eigenstates of the full Hamiltonian, that couple to
  the repulsive state due to the anharmonic trapping potential which
  manifests in avoided crossings, see \figref{fig:full_spec}. }.
The results strongly imply that the inelastic resonances are the
dominant cause for the appearance of the two distinct loss features in
the experiment by E.\ Haller et al.\ \cite{cold:hall10b} as was
already suggested in \cite{cold:sala12}. In general, the effect of
COM-REL motion coupling can have a significant impact on the stability
of 1D quantum gases and should therefore be considered in current 1D
experiments.

\acknowledgments{The authors gratefully acknowledge support from the
  \textit{Else-Neumann Stiftung}, \textit{Studienstiftung des deutschen
    Volkes}, \textit{Fonds der Chemischen Industrie}, IMPRS-QD,
  Helmholtz Alliance HA216/EMMI, the Heidelberg Center for Quantum
  Dynamics and ERC Starting Grant 279697. We thank M.~G\"arttner for
  helpful discussions.}

%

\clearpage
\newpage

\section{Supplemental Material}

\subsection{Measurement of the single particle excitation frequencies in the trap}  
We excite individual atoms in the ground state of the trap to higher
trap states by periodically modulating the center position or the
confinement length of the trap with a certain frequency $\omega$. The
transition frequencies are determined from fits to the excitation
spectra (see \cite{cold:zurn_phd}). The results are given in Table
\ref{tab:trap-frequencies}

\begin{table} [h!]
  \begin{center}
    \begin{tabular}{|c | c | c|  }
      \hline
        h.o. number  & frequency 										  \\ 
       ($n_x$,$n_y$,$n_z$)& $\omega/2\pi\;$ [kHz] \\ \hline
    	 (0,0,1) &1.486(0.011)\\ 
       (0,0,2) &2.985(0.010)\\ 
       (0,0,4) &2.897(0.020)\\ \hline
			 (1,0,0) &13.96(0.08) \\ 
       (0,1,0) &14.82(0.09) \\ 
       (2,0,0) &26.43(0.27) \\ 
       (0,2,0) &28.26(0.25) \\ \hline
    \end{tabular}
  \end{center}
  \caption{Transition frequencies in the trap of atoms excited in the trap. We denote the transitions by the corresponding quantum number of a harmonic oscillator. }
  \label{tab:trap-frequencies}
\end{table}

\subsection{Determination of the elastic CIR}

To determine the elastic CIR we only consider one resonance following
\cite{cold:peng10, cold:zhan11} although we have found an anisotropy
in our system. We calculate the position of the elastic CIR from the
transversal confinement length
\begin{equation}
d_{\perp}=\sqrt{\hbar/\mu \omega_{\perp}}\, .
\label{eq:osz length}
\end{equation}
To determine the mean trap frequency $\omega_{\perp}$ we fit the
Gaussian shape of the optical beam which creates the trapping
potential to the measured transition frequencies (see
\cite{cold:zurn_phd}). From the two profiles in radial direction we
deduce the corresponding transition frequencies between the lowest
single particle bound states in the gaussian potential and calculate
the mean frequency
\begin{equation}
\omega_{\perp}=\frac{1}{2}\,(\omega_{(1,0,0)\text{fit}}+\omega_{(0,1,0)\text{fit}})_ =  2\pi\times (14.22\pm 0.35)\, \text{kHz}\,.
\end{equation}
Using equation \ref{eq:osz length} as the input of the theory of
\cite{cold:berg03} and the calibration a(B) of \cite{cold:zurn12b} we
calculate the position of the elastic CIR.  The propagated error of
the mean trap frequency serves as the systematic uncertainty of the
position of the CIR.

\subsection{Fitting the trapping potential}

In an ideal and simplified case, the shape of the external potential
is determined by the profile of a Gaussian beam which results in the
potential
\begin{align}
  V(\mathbf{r}) = 
  V_0 \left( \frac{w_0}{w(z)} \right)^2 \exp \left(
    \frac{-2(x^2+y^2)}{w^2(z)} \right)
  \label{eq:gauss_beam}
\end{align}
where $ w(z) = w_0 \, \sqrt{ 1+ {\left( \frac{z}{z_\mathrm{R}}
    \right)}^2 }$ and $w_0 = w(0)$ is the waist size of the beam and $
z_\mathrm{R} = \frac{\pi w_0^2}{\lambda} $ is the Rayleigh range. The
measurement of the excitation energies, Table
\ref{tab:trap-frequencies}, reveals an ellipticity of
$\omega_{x\,2-0}/\omega_{y\,2-0}=1.07$ in radial direction which
cannot be justified by the beam potential
\eqref{eq:gauss_beam}. Hence, for the theoretical description, the
potential \eqref{eq:gauss_beam} is expanded in a Taylor series around
the origin up to second order which results in a harmonic potential of
the form
\begin{align}
  V(\mathbf{r}) = \frac{\hbar}{2}\ \sum_{j=x,y,z}\omega_j^2\, j^2.
\end{align}
To address the ellipticity in the radial direction, different harmonic
$\omega_j$ frequencies for the $x$ and $y$ direction are chosen. In
the harmonic approximation, however, the coupling between COM and REL
degrees of freedom vanishes. This is surly not the case for the
finite, anharmonic experimental potential. Hence, for the model
potential higher order terms are introduced to include COM-REL
coupling. It has been demonstrated that sextic potentials,
i.e. expansions of a $\sin^2$ optical lattice potential up to order
six
\begin{align}
  V(\mathbf{r}) = \sum_{j=x,y,z} \frac{2}{45} V_{j} k_{j}^{6} j^{6} -
  \frac{1}{3} V_{j} k_{j}^{4} j^{4} + V_{j} k_{j}^{2} j^{2} ,
  \label{eq:sextic_absolute}
\end{align}
are well suited to describe single well potentials including COM-REL
coupling \cite{cold:gris09}.\\ 
The resulting fitting parameters (given in atomic units) are $k_x = 9.23 \times 10^{-5}$ a.u.,
$k_y = 8.77 \times 10^{-5}$ a.u., $k_z = 1.58 \times 10^{-5}$ a.u.,
$V_x = 6.82 \times 10^{-12}$ a.u., $V_y = 6.58 \times 10^{-12}$ a.u.\ and $V_z = 2.33 \times 10^{-12}$ a.u.

\subsection{COM-REL coupling involving the state $\mathbf{\ket{\psi^{(b)} \, \Phi_{2,0,0}}}$} 

\begin{figure} [th!]
\centering
\includegraphics [width=0.45\textwidth] {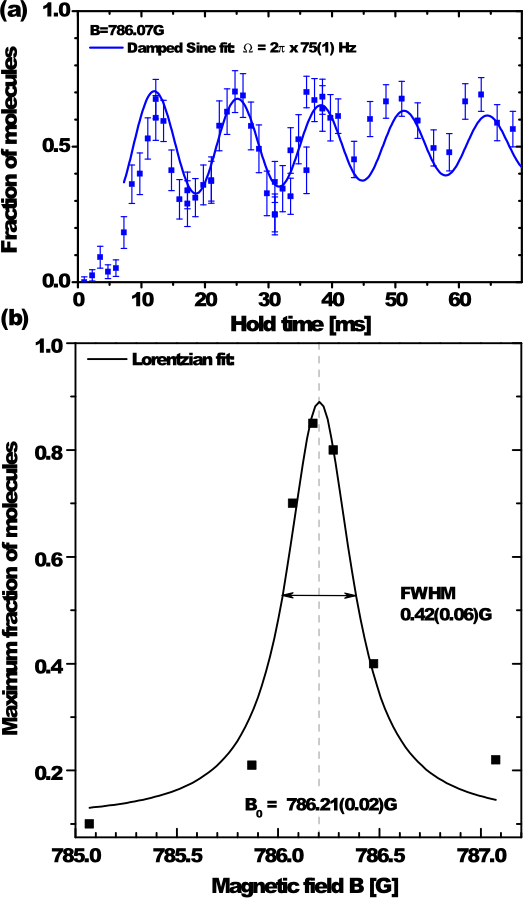}
	\caption{Dynamic of the COM-REL motion coupling. (a) Oscillation between the non-bound state and the COM excited molecular state $\ket{\psi^{(b)} \, \Phi_{2,0,0}}$. (b) Maximum fraction of molecules extracted from measurements similar to a) at different magnetic fields. 
	}
	\label{fig:supplemental-fig1} 
\end{figure}

The measurement of coherent coupling between the repulsive non-bound
state and the COM-excited bound state $\ket{\psi^{(b)} \,
  \Phi_{2,0,0}}$ has been performed in a similar way as the one
presented in the main text with the only difference that an additional
linear potential had been applied in z-direction (see
Fig.\ref{fig:supplemental-fig1} ).
 The additional potential reads
\begin{equation}
V_{\text{lin}}= \mu_m B' z
\end{equation}
where $\mu_m$ is the Bohr magneton and $B'=18.92\,$G/cm a magnetic
field gradient. The additional potential slightly shifts the energy of
the non-bound repulsive state $\ket{\psi_{(0)} \, \Phi_{0,0,0}}$ and
thus also shifts the position of the COM-REL coupling resonances. Yet,
the influence on the coupling strength should be negligible. Hence we
can use the results for the Rabi frequency and the width presented in
Fig. \ref{fig:supplemental-fig1} for a comparison with the numerical
calculation which has been performed without an additional potential
in z-direction.

\end{document}